\documentclass[letterpaper,11pt]{article}

\usepackage{amsthm}
\usepackage{amssymb}
\usepackage{amsmath}
\usepackage{amsfonts}
\usepackage{amstext}

\usepackage[sort,numbers]{natbib}

\usepackage{fullpage}

\usepackage{color}

\usepackage[T1]{fontenc}
\usepackage[utopia]{mathdesign}

\newcommand{\tinyspace}{\mspace{1mu}}

\newcommand*{\bra}[1]{\langle #1|}
\newcommand*{\ket}[1]{|#1\rangle}
\newcommand{\ketbra}[1]{\ket{#1}\bra{#1}}

\DeclareRobustCommand\openone{\leavevmode\hbox{\upshape\small1\normalsize\kern-.33em1}}%
\newcommand{\tensor}{\otimes}
\newcommand{\tprod}{\otimes}

\newcommand{\norm}[1]{\left\lVert\tinyspace#1\tinyspace\right\rVert}
\newcommand{\tnorm}[1]{\norm{#1}_{\mathrm{tr}}}
\newcommand{\dnorm}[1]{\norm{#1}_{\diamond}}

\newcommand{\dm}[1]{\dim \mathcal{#1}}

\newcommand{\linear}[1]{\mathbf{L}(\mathcal{#1})}
\newcommand{\density}[1]{\mathbf{D}(\mathcal{#1})}

\newcommand{\transform}[1]{\mathbf{T}(\mathcal{#1})}

\newcommand{\tidentity}[1]{I_{\mathcal{#1}}}

\newcommand{\tr}{\operatorname{tr}}

\newcommand{\ptr}[1]{\tr_\mathcal{#1}}

\newcommand{\prob}[1]{\textup{\textsc{#1}}}
\newcommand{\class}[1]{\textup{\textsf{#1}}}

\newtheorem{theorem}{Theorem}
\newtheorem{lemma}[theorem]{Lemma}

\theoremstyle{definition}

\newtheorem{problem}[theorem]{Problem}

\usepackage{hyperref}

\title{Computational distinguishability of degradable and antidegradable channels}

\author{Bill Rosgen \\
  Centre for Quantum Technologies \\
  National University of Singapore}

\date{11 November, 2009}

\begin{document}

\maketitle

\begin{abstract}
  A channel is degradable if there exists a second channel that maps
  the output state of the channel to the environment state.  These 
  channels satisfy the property that the output state contains
  more information about the input than the environment does.  A
  complementary class of channels is the antidegradable channels,
  which admit channels that map the environment state to the output
  state of the channel.
  In this paper we show that the computational problem of
  distinguishing two channels remains \class{PSPACE}-complete when
  restricted to these classes of channels.  This is shown using a
  construction of Cubitt, Ruskai, and Smith~\cite{CubittR+08} that
  embeds any channel into a degradable channel, and a related
  construction for the case of antidegradable channels.
\end{abstract}


\section{Introduction}\label{chap-degr}

The task of distinguishing two quantum channels is one of the most
fundamental problems in quantum information.  This problem, a weak
form of process tomography, asks not to completely characterize an unknown
quantum channel, but to identify it as one of two known channels.
An equivalent formulation asks if there is an input
state on which two known channels produce distinct output states.
When this problem is phrased computationally, with the two known
channels specified as quantum circuits, the resulting problem is
complete for the class \class{PSPACE}~\cite{RosgenW05}, the
class of all problems that can be solved in a polynomially bounded
amount of space.

In light of this hardness, it is natural to ask if restricted versions
of the problem are less difficult.  In the case that the channels to
be distinguished are unitary, the problem is ``only''
\class{QMA}-complete~\cite{JanzingW+05} (see also~\cite{JiW09}), where 
\class{QMA} is the class of all those problems that can be efficiently
verified with a quantum computer.  In the case that the input circuits
implement convex mixtures of unitary 
operation~\cite{Rosgen08additivity} or have short circuit
descriptions~\cite{Rosgen08distinguishing} the problem is known to
remain \class{PSPACE}-complete.  In this paper we add two more
restricted classes of channels to this list: the degradable channels
and the antidegradable channels.

A quantum channel $\Phi$ is degradable if there exists a second
channel that maps the output of $\Phi$ to the state of the environment
after applying $\Phi$.  More precisely, a channel $\Phi \colon
\linear{A} \to \linear{B}$ that is given by
\[ \Phi(\rho) = \ptr{E} U( \rho \tprod \ket 0 \bra 0) U^*, \] is
called degradable if there exists a channel $\Delta_\Phi \colon
\linear{B} \to \linear{E}$ such that
\[ (\Delta_\Phi \circ \Phi)(\rho) = \ptr{B} U( \rho \tprod \ket 0 \bra 0)
U^* = \Phi^C(\rho). \]
The channel $\Phi^C$ is called the complementary channel to $\Phi$,
and it is only defined up to an isometry, since it depends on the
Stinespring representation of $\Phi$.  This does not affect the notion
of degradability, however, as this isometry can be incorporated into
the degrading map.
These channels were introduced by Shor and
Devetak~\cite{DevetakS05} to study the capacity of a channel for
transmitting quantum information.  Notice that the set of degradable channels
is not convex: any unitary channel is degradable, but the completely
depolarizing channel is not, and it can be written as a convex
combination of unitary channels (see~\cite{BoykinR03} for an example
of such a construction).

A channel is called antidegradable if the complementary channel is degradable.
Alternately, a channel is $\Phi$ antidegradable is there exists a map
$A_\Phi$ such that $A_\Phi \circ \Phi^C = \Phi$, where once again the channel
$\Phi^C$ is only defined up to an isometry, but this isometry can also
be part of the map $A_\Phi$, so that the antidegradable channels are also
well-defined.  This class of channels was introduced by Wolf and
P{\'e}rez-Garcia~\cite{WolfP07}.  These channels can be informally
thought of as the very noisy channels that lose more information to the
environment than they preserve in the output.
A thorough discussion of the degradable and antidegradable
channels can be found in~\cite{CubittR+08}, where it is shown that,
unlike the degradable channels, the set of antidegradable channels is convex.

The degradable and antidegradable channels are interesting from a
quantum information perspective.  A no-cloning argument implies
that the antidegradable channels have zero capacity for the
transmission of quantum information~\cite{GiovannettiF05}.
It is also known that the coherent information is additive on
degradable channels, which implies that the quantum capacity is given by
the coherent information of a single use of the channel, i.e.\ the
formula for the quantum capacity does not require
regularization~\cite{DevetakS05}.

As the degradable and antidegradable channels have nice properties
with respect to the transmission of quantum information, it might be
hoped that similar properties extend to the transmission of classical
information.  In the case of the Holevo (or $\chi$-)capacity, it is shown
in~\cite{CubittR+08} that the additivity of this quantity on
degradable channels is equivalent to the general case, making use of a
result from~\cite{FukudaW07}.  As it is also known that this
additivity problem is equivalent on the complementary class of
channels~\cite{Holevo07, KingM+07}, this implies that the additivity
of the antidegradable channels is also equivalent to the general case.
Finally, using the recent result of Hastings~\cite{Hastings09}, there
are degradable and antidegradable channels that do not have additive
Holevo capacity.

Interestingly, we can adapt the construction used by Cubitt, Ruskai,
and Smith~\cite{CubittR+08} to show that the quantum circuit
distinguishability problem restricted to either the degradable or
antidegradable channels remains \class{PSPACE}-complete.  These results
are the focus of this paper.

The complexity class \class{PSPACE} is the class of all problems that
are solvable on a classical computer in a polynomially bounded amount
of space.  A recent result of Jain et al. shows that this class is
equal to \class{QIP}, the class of all problems that can be
interactively verified~\cite{JainJ+09}.  This implies that the problem
of distinguishing degradable (or antidegradable) quantum channels
exactly captures the difficulty of classical space bounded
computation.  It is known that \class{PSPACE} contains the class
\class{QMA}, corresponding to quantum one-round verifiable
computation.  The class \class{QMA} in turn contains the class
\class{BQP}, which captures the notion of efficient quantum
computation.  It is believed that both of these containments are
strict, though proving this for either of them would be a
major breakthrough in complexity theory.

The remainder of the paper is organized as follows.
Section~\ref{scn-prelim} presents some notation and results that will
be used in the rest of the paper.  Section~\ref{scn-sim} presents the
construction due to Cubitt et al.~\cite{CubittR+08} that embeds any
channel into a degradable channel, as well as a related construction
that embeds an arbitrary channel into an antidegradable channel.  The
main result of the paper is contained in Section~\ref{scn-dist}, where
these two constructions are used to show that the problems of
distinguishing the degradable channels and the antidegradable channels
remain \class{PSPACE}-complete.

\section{Preliminaries}\label{scn-prelim}

Here we introduce some technical concepts used
throughout the paper.  The notation used is standard: the
reader familiar with quantum information is invited to skim this section.

Throughout the paper scripted capital letters $\mathcal{A,B,C,
  \ldots}$ will refer to Hilbert spaces, all of which will be finite
dimensional.  The set of all linear operators mapping $\mathcal{A}$ to
$\mathcal{B}$ is denoted $\linear{A,B}$.  The set of quantum states, or
density matrices, on a space $\mathcal{A}$ is $\density{A}$: these are
simply the positive semidefinite operators in $\linear{A,A}$ with unit
trace.

A quantum channel is simply a map that takes density operators to
density operators, even when applied to part of a larger system.
These are given by the completely positive and trace preserving maps
from $\linear{A}$ to $\linear{B}$.  These maps are represented using
capital Greek letters $\Phi, \Psi, \ldots$, and the set of all such
maps is given by $\transform{A,B}$.  The notation $\tidentity{A}$
refers to the identity map on $\linear{A}$.

Given one use of an unknown channel $\Phi \in \transform{A,B}$, that is promised to be one
of two known channels $\Phi_1$ and $\Phi_2$, what is the maximum
probability that the channel can be identified?  This quantity is
given by the \emph{diamond norm} of $\Phi_1 - \Phi_2$, which is
\begin{equation}\label{eqn-dnorm}
  \dnorm{\Phi_1 - \Phi_2} 
  = \max_{\rho} \tnorm{ 
    (\Phi_1 \tprod \tidentity{F})(\rho) - (\Phi_2 \tprod \tidentity{F})(\rho)},
\end{equation}
where the space $\mathcal{F}$ is a space of the same dimension as
$\mathcal{A}$ and the maximization is taken
over all density matrices in $\density{A \tprod F}$.
The maximum probability that an unknown operation in $\{\Phi_1,
\Phi_2\}$ can be correctly identified with a single use is given by
\begin{equation*}
  \frac{1}{2} + \frac{1}{4} \dnorm{\Phi_1 - \Phi_2},
\end{equation*}
which is one reason that this norm is central to the study of channel distinguishability.
The diamond norm can be more generally defined over any linear operator
mapping $\linear{A}$ to $\linear{B}$ (see~\cite{KitaevS+02} for such a
definition, as well as some properties of this norm).  The fact that
we may restrict the maximum in Equation~\eqref{eqn-dnorm} to a density
matrix in the case of the difference of two completely positive maps
can be found in~\cite{RosgenW05}.
This norm is closely related to the \emph{completely bounded norm}, in
fact, $\dnorm{\Phi} = \norm{\Phi^*}_{\mathrm{cb}}$, where $\Phi^*$ is
the adjoint of $\Phi$ with respect to the Hilbert-Schmidt inner product.

One important property of the diamond norm of two channels is that
applying it to several copies increases the norm.  Intuitively this is
obvious: given $k$ copies of an unknown channel it is expected that it
is easier to identify.  The diamond norm of $\Phi_1^{\tprod k} -
\Phi_2^{\tprod k}$ corresponds to a non-adaptive strategy with access
to $k$ copies of the two channels.  The following Lemma
from~\cite{RosgenW05}, which appears there as part of an efficient
polarization procedure for the diamond norm, gives simple bounds on
the norm as the number of copies increases.
\begin{lemma}\label{lem-direct-product}
  Let $\Phi_1, \Phi_2 \in \transform{A,B}$ have
  $\dnorm{\Phi_1 - \Phi_2} = \delta > 0$.  Then for any positive integer
  $k$
  \begin{equation*}
    2 - 2e^{ \frac{-k \delta^2}{8} }
    < \dnorm{ \Phi_1^{\tprod k} - \Phi_2^{\tprod k}}
    \leq  k \delta.
  \end{equation*}
\end{lemma}
\noindent This lemma will be used to show that parallel repetition reduces
the error introduced by the reduction of the distinguishability
problem to the cases of degradable and antidegradable channels.

In order to capture the difficulty of distinguishing implementations
that represent efficient quantum computation, the input channels to
the computational problems studied here are given as mixed-state
circuits.  This model, proposed by Aharonov et al.~\cite{AharonovK+98},
consists of circuits in the usual unitary model, with two additional
gates.  These two gates are the introduction of fresh ancillary qubits
in the $\ket 0$ state and the partial trace of a qubit, both of which
are non-unitary operations.  The resulting circuit model allows for
the (approximate) implementation of any quantum channel.

Notice that any circuit in this model can be efficiently converted
into a circuit that first introduces any ancillary qubits, then
performs a unitary circuit, and finally traces out any qubits that are
not part of the output.  This is due to the fact that introducing
qubits earlier and tracing out qubits later does not affect the rest
of the circuit.  Thus, any mixed-state circuit can be assumed to be in
the form of a Stinespring dilation.  This property will be essential
to the construction of degradable and antidegradable simulations in Section~\ref{scn-sim}.

Channels are represented using mixed-state circuits as this
provides a succinct representation of efficient quantum algorithms.
Representing channels using a (potentially) exponentially larger
representation, such as a Kraus decomposition, renders the problem
solvable in classical polynomial time~\cite{Watrous09semidefinite,
  Ben-AroyaT-S09}, but in this model we lose the connection to
efficient quantum algorithms as the descriptions of the channels are
of size exponential in the number of input and output qubits.  To
maintain relevance in the case of practical distinguishability, such
as between two physical implementations of quantum algorithms, we need
an input description that scales logarithmically in the Hilbert space
dimension, and so we use the mixed-state circuit model.

\section{Simulations of channels}\label{scn-sim}

In this section we present two related constructions: the first embeds any
channel into a degradable channel and the second embeds any channel
into an antidegradable channel.  Efficient quantum circuits for these
problems (as well as the corresponding degrading and antidegrading
maps) are presented.

\subsection{Degradable channels}\label{scn-degr-sim}

Given a quantum channel $\Phi$ we seek to simulate $\Phi$ by a
degradable channel $\Psi$ that has similar properties with respect to
distinguishability.  This can be done by adapting the construction
used by Cubitt, Ruskai, and Smith~\cite{CubittR+08} for the case of
the minimum output entropy.

To describe this construction, we assume that $\Phi \in
\transform{A,B}$ with $\dm{A} = \dm{B}$, i.e.\ that
the original channel has identical input and output dimension.  This
assumption can be made without loss in generality by padding the smaller space with
unused qubits, since these qubits will not affect the diamond
norm used to define distinguishability.  As the spaces $\mathcal{A}$
and $\mathcal{B}$ have the same dimension, they are isomorphic, and so
we may view $\Phi$ as a channel in $\transform{A,A}$.

The channel $\Phi$, given as input to a computational problem, is
specified as a mixed-state quantum circuit.  Such a circuit, as
discussed in Section~\ref{scn-prelim}, can be efficiently transformed
into one that first introduces any ancillary qubits, then performs
some unitary circuit, and finally traces out any qubits that are not
part of the output state.  To this end, let the circuit $\Phi \in
\transform{A,A}$ use the space $\mathcal{E}$ for ancillary qubits, and
let $U$ be the unitary that is applied to the space $\mathcal{A \tprod
  E}$.  Stated formally, the channel $\Phi$ is specified by 
\begin{equation}\label{eqn-degr-stinespring}
  \Phi(\rho) = \ptr{E} U (\rho \tprod \ketbra 0) U^*.
\end{equation}
This notation will be used to construct
the degradable simulation of $\Phi$.

\begin{figure}
  \begin{center}
\setlength{\unitlength}{3947sp}%
\begingroup\makeatletter\ifx\SetFigFont\undefined%
\gdef\SetFigFont#1#2#3#4#5{%
  \reset@font\fontsize{#1}{#2pt}%
  \fontfamily{#3}\fontseries{#4}\fontshape{#5}%
  \selectfont}%
\fi\endgroup%
\begin{picture}(2877,2124)(-164,-1648)
\put(-149,-1036){\makebox(0,0)[lb]{\smash{{\SetFigFont{12}{14.4}{\rmdefault}{\mddefault}{\updefault}{\color[rgb]{0,0,0}$\ket 0$}%
}}}}
\thinlines
{\color[rgb]{0,0,0}\put(-149,-211){\line( 1, 0){1350}}
}%
{\color[rgb]{0,0,0}\put(-149,-286){\line( 1, 0){1350}}
}%
{\color[rgb]{0,0,0}\put(-149,-361){\line( 1, 0){1350}}
}%
{\color[rgb]{0,0,0}\put(-149,-436){\line( 1, 0){1350}}
}%
{\color[rgb]{0,0,0}\put(-149,-136){\line( 1, 0){1350}}
}%
{\color[rgb]{0,0,0}\put(2101,239){\circle*{76}}
}%
{\color[rgb]{0,0,0}\put(2701,-211){\line(-1, 0){900}}
}%
{\color[rgb]{0,0,0}\put(2701,-286){\line(-1, 0){900}}
}%
{\color[rgb]{0,0,0}\put(2701,-361){\line(-1, 0){900}}
}%
{\color[rgb]{0,0,0}\put(2701,-436){\line(-1, 0){900}}
}%
{\color[rgb]{0,0,0}\put(2701,-136){\line(-1, 0){900}}
}%
{\color[rgb]{0,0,0}\put(2401,-961){\line(-1, 0){600}}
}%
{\color[rgb]{0,0,0}\put(2401,-1036){\line(-1, 0){600}}
}%
{\color[rgb]{0,0,0}\put(2401,-1111){\line(-1, 0){600}}
}%
{\color[rgb]{0,0,0}\put(2401,-1186){\line(-1, 0){600}}
}%
{\color[rgb]{0,0,0}\put(2401,-886){\line(-1, 0){600}}
}%
{\color[rgb]{0,0,0}\put(2401,-811){\line(-1, 0){600}}
}%
{\color[rgb]{0,0,0}\put(151,-961){\line( 1, 0){1050}}
}%
{\color[rgb]{0,0,0}\put(151,-1036){\line( 1, 0){1050}}
}%
{\color[rgb]{0,0,0}\put(151,-1111){\line( 1, 0){1050}}
}%
{\color[rgb]{0,0,0}\put(151,-1186){\line( 1, 0){1050}}
}%
{\color[rgb]{0,0,0}\put(151,-886){\line( 1, 0){1050}}
}%
{\color[rgb]{0,0,0}\put(151,-811){\line( 1, 0){1050}}
}%
{\color[rgb]{0,0,0}\put(2101,-1411){\circle{76}}
}%
{\color[rgb]{0,0,0}\put(901,239){\line( 1, 0){1800}}
}%
{\color[rgb]{0,0,0}\put(1501,239){\line( 0,-1){300}}
}%
{\color[rgb]{0,0,0}\put(451, 14){\framebox(450,450){$H$}}
}%
{\color[rgb]{0,0,0}\put(151,239){\line( 1, 0){300}}
}%
{\color[rgb]{0,0,0}\put(1201,-1261){\framebox(600,1200){$U$}}
}%
{\color[rgb]{0,0,0}\put(151,-1411){\line( 1, 0){2250}}
}%
{\color[rgb]{0,0,0}\put(2401,-811){\vector( 0,-1){825}}
}%
{\color[rgb]{0,0,0}\put(2101,239){\line( 0,-1){1688}}
}%
{\color[rgb]{0,0,0}\put(-149,-511){\line( 1, 0){1350}}
}%
{\color[rgb]{0,0,0}\put(1801,-511){\line( 1, 0){900}}
}%
\put(-149,164){\makebox(0,0)[lb]{\smash{{\SetFigFont{12}{14.4}{\rmdefault}{\mddefault}{\updefault}{\color[rgb]{0,0,0}$\ket 0$}%
}}}}
\put(-149,-1486){\makebox(0,0)[lb]{\smash{{\SetFigFont{12}{14.4}{\rmdefault}{\mddefault}{\updefault}{\color[rgb]{0,0,0}$\ket 0$}%
}}}}
{\color[rgb]{0,0,0}\put(1501,239){\circle*{76}}
}%
\end{picture}%
 \end{center}
  \caption[Reduction to a degradable channel]{The degradable channel
    $\Psi$ constructed from $\Phi$, where $U$ is the unitary
    given in Equation~\eqref{eqn-degr-stinespring}.}
  \label{fig-degrad-reduction}
\end{figure}
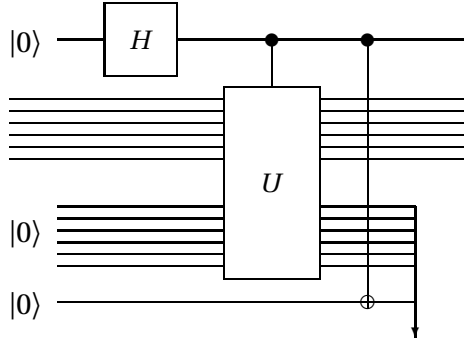

The idea is to implement the channel
\begin{equation}\label{eqn-degr-const}
  \Psi(\rho) 
  = \frac{1}{2} \ketbra 0 \tprod \rho + \frac{1}{2} \ketbra 1 \tprod \Phi(\rho),
\end{equation}
which has been used by Cubitt, Ruskai, and Smith~\cite{CubittR+08} to
reduce the (non-)additivity of the minimum output entropy to the
degradable case.
This is the channel that applies $\Phi$ with probability
$1/2$, does nothing to the input with probability $1/2$, and leaves a
flag on the output to indicate which case has occurred.  The channel
$\Psi$ maps states on $\mathcal{A}$ to mixed states on $\mathcal{C
  \tprod A}$, where $\mathcal{C}$ is the space of dimension two
corresponding to flag state.  Using the implementation of $\Phi$ in
Equation~\eqref{eqn-degr-stinespring}, a circuit for the channel
$\Psi$ is given in Figure~\ref{fig-degrad-reduction}.
The idea in this implementation is that the top ancillary qubit,
corresponding do the space $\mathcal{C}$, is placed in the $\ket +$
state, which results in the circuit for $\Phi$ being applied with
probability one-half.  This control qubit is then `copied' onto one of
the environment qubits, so that the resulting output state is the
mixture given in Equation~\eqref{eqn-degr-const}.

To see that $\Psi$ is a degradable channel, we construct
the map that takes the output state of $\Psi$ to the state of the
environment.  The construction of the channel $\Psi$, as well as a proof that it is
degradable can be found in~\cite{CubittR+08}.  This proof is quite
simple, and so it is repeated here.
Before we construct the degrading map, however, notice that the
complementary channel of $\Psi \in \transform{A,C \tprod A}$, which is given by reversing
the output and environment spaces, is
\begin{equation}\label{eqn-degrad-compl}
  \Psi^C(\rho) 
  = \frac{1}{2} \ketbra 0 \tprod \ketbra 0 + \frac{1}{2} \ketbra 1
  \tprod \Phi^C(\rho),
\end{equation}
where the channel $\Phi^C \in \transform{A,E}$ is the complement of
the original channel $\Phi \in \transform{A,A}$, given by
\[ \Phi^C(\rho) = \ptr{A} U ( \rho \tprod \ket 0 \bra 0) U^*. \] 
These complementary channels are only defined up to an isometry, since a Stinespring
dilation is defined only up to an isometry on the environment space,
but for the present purpose, \emph{any} complementary channel suffices.

\begin{figure}
  \begin{center}
\setlength{\unitlength}{3947sp}%
\begingroup\makeatletter\ifx\SetFigFont\undefined%
\gdef\SetFigFont#1#2#3#4#5{%
  \reset@font\fontsize{#1}{#2pt}%
  \fontfamily{#3}\fontseries{#4}\fontshape{#5}%
  \selectfont}%
\fi\endgroup%
\begin{picture}(2577,1749)(1186,-1273)
\put(1201,-1036){\makebox(0,0)[lb]{\smash{{\SetFigFont{12}{14.4}{\rmdefault}{\mddefault}{\updefault}{\color[rgb]{0,0,0}$\ket 0$}%
}}}}
\thinlines
{\color[rgb]{0,0,0}\put(1501,-1036){\line( 1, 0){1050}}
}%
{\color[rgb]{0,0,0}\put(1501,-1111){\line( 1, 0){1050}}
}%
{\color[rgb]{0,0,0}\put(1501,-1186){\line( 1, 0){1050}}
}%
{\color[rgb]{0,0,0}\put(1501,-886){\line( 1, 0){1050}}
}%
{\color[rgb]{0,0,0}\put(1501,-811){\line( 1, 0){1050}}
}%
{\color[rgb]{0,0,0}\put(2851,239){\circle*{76}}
}%
{\color[rgb]{0,0,0}\put(3151,-961){\line( 1, 0){600}}
}%
{\color[rgb]{0,0,0}\put(3151,-1036){\line( 1, 0){600}}
}%
{\color[rgb]{0,0,0}\put(3151,-1111){\line( 1, 0){600}}
}%
{\color[rgb]{0,0,0}\put(3151,-1186){\line( 1, 0){600}}
}%
{\color[rgb]{0,0,0}\put(3151,-886){\line( 1, 0){600}}
}%
{\color[rgb]{0,0,0}\put(3151,-811){\line( 1, 0){600}}
}%
{\color[rgb]{0,0,0}\put(1201,-511){\line( 1, 0){1350}}
}%
{\color[rgb]{0,0,0}\put(1201,-211){\line( 1, 0){1350}}
}%
{\color[rgb]{0,0,0}\put(1201,-286){\line( 1, 0){1350}}
}%
{\color[rgb]{0,0,0}\put(1201,-361){\line( 1, 0){1350}}
}%
{\color[rgb]{0,0,0}\put(1201,-436){\line( 1, 0){1350}}
}%
{\color[rgb]{0,0,0}\put(1201,-136){\line( 1, 0){1350}}
}%
{\color[rgb]{0,0,0}\put(3151,-286){\line( 1, 0){300}}
}%
{\color[rgb]{0,0,0}\put(3151,-361){\line( 1, 0){300}}
}%
{\color[rgb]{0,0,0}\put(3151,-436){\line( 1, 0){300}}
}%
{\color[rgb]{0,0,0}\put(3151,-511){\line( 1, 0){300}}
}%
{\color[rgb]{0,0,0}\put(3151,-211){\line( 1, 0){300}}
}%
{\color[rgb]{0,0,0}\put(3151,-136){\line( 1, 0){300}}
}%
{\color[rgb]{0,0,0}\put(2551,-1261){\framebox(600,1200){$U$}}
}%
{\color[rgb]{0,0,0}\put(1801, 14){\framebox(450,450){$X$}}
}%
{\color[rgb]{0,0,0}\put(1201,239){\line( 1, 0){600}}
}%
{\color[rgb]{0,0,0}\put(2851,-61){\line( 0, 1){300}}
}%
{\color[rgb]{0,0,0}\put(2251,239){\line( 1, 0){1500}}
}%
{\color[rgb]{0,0,0}\put(3451,-136){\vector( 0,-1){525}}
}%
{\color[rgb]{0,0,0}\put(1501,-961){\line( 1, 0){1050}}
}%
\end{picture}%
 \end{center}
  \caption[Degrading channel for the channel in
  Figure~\ref{fig-degrad-reduction}]{The degrading channel $\Delta_\Psi$
    corresponding to the channel in $\Psi$ in
    Figure~\ref{fig-degrad-reduction}.}
  \label{fig-degrad-degrading}
\end{figure}
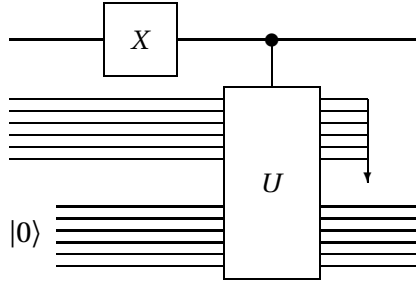

Given Equation~\eqref{eqn-degrad-compl}, it is not hard to construct
the degrading map $\Delta_\Psi$.
Starting with the output of $\Psi$
\begin{equation*} 
  \Psi(\rho) 
  = \frac{1}{2} \ket 0 \bra 0 \tprod \rho + \frac{1}{2} \ket 1 \bra 1 \tprod \Phi(\rho),
\end{equation*}
as given by Equation~\eqref{eqn-degr-const},
this channel can, based on a measurement of the flag state in the space $\mathcal{C}$
output one of the two states $\ket 0 \bra 0$ and $\Phi^C(\rho)$.
More formally, when the flag state is $\ket 0$ the state in $\mathcal{A}$ is the original
input $\rho$, so the channel can apply $\Phi^C$ to produce
$\Phi^C(\rho)$.
On the other hand, when this flag state is $\ket
1$, the degrading map outputs $\ket 0 \bra 0$, which can be
done by producing the correct number of untouched ancillary qubits as
output.  All that remains in to invert the flag qubit to get exactly
the output of $\Psi^C$.  A circuit implementation of the channel $\Delta_\Psi$ is
presented in Figure~\ref{fig-degrad-degrading}.  The fact that this
channel has an efficient circuit implementation is not important for
the main result: this is merely a simple way to specify the channel.
We can verify that this map performs the required operation by observing that
\begin{align*}
  \Delta_\Psi( \Psi(\rho) )
  &= \frac{1}{2} \Delta_\Psi\left(\ket 0 \bra 0 \tprod \rho + \ket 1 \bra 1 \tprod \Phi(\rho)\right) \\
  &= \frac{1}{2} \ket 1 \bra 1 \tprod \Phi^C(\rho) + \frac{1}{2} \ket 0 \bra 0 \tprod \ket 0 \bra 0 \\
  &= \Psi^C(\rho),
\end{align*}
where the final equality is Equation~\eqref{eqn-degrad-compl}.
This argument, due to Cubitt, Ruskai, and Smith~\cite{CubittR+08}
proves that the channel $\Psi$ is degradable.  In the next section we
adapt this construction to the case of the antidegradable channels.

\subsection{Antidegradable channels}\label{scn-adegr-sim}

In this section a construction very similar to that used in
Section~\ref{scn-degr-sim} is presented that takes any circuit
$\Phi$ to a circuit $\Psi$ implementing an antidegradable channel.  The idea
is to (with probability one-half) send the input state to the
environment, so that the channel that maps the environment state to
the output state will have a copy of the input state.  This
construction (and the proof that it produces an antidegradable
channel) is very similar to the construction used for
degradable channels.

Once again we may assume that $\Phi$ implements a channel in
$\transform{A,A}$, i.e.\ that $\Phi$ has the same input and output
dimension, by embedding the smaller space into the larger, if
necessary.  As in Section~\ref{scn-degr-sim}, the constructed
circuit $\Psi$ will use one additional output qubit, implementing an
antidegradable transformation in $\transform{A, C \tprod A}$.

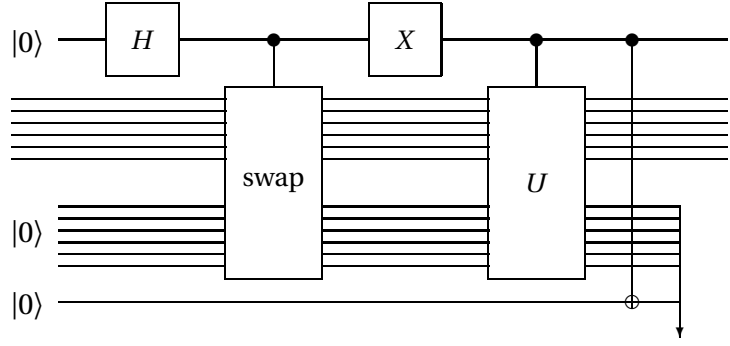
\begin{figure}
  \begin{center}
\setlength{\unitlength}{3947sp}%
\begingroup\makeatletter\ifx\SetFigFont\undefined%
\gdef\SetFigFont#1#2#3#4#5{%
  \reset@font\fontsize{#1}{#2pt}%
  \fontfamily{#3}\fontseries{#4}\fontshape{#5}%
  \selectfont}%
\fi\endgroup%
\begin{picture}(4527,2124)(-464,-1648)
\put(-449,-1036){\makebox(0,0)[lb]{\smash{{\SetFigFont{12}{14.4}{\rmdefault}{\mddefault}{\updefault}{\color[rgb]{0,0,0}$\ket 0$}%
}}}}
\thinlines
{\color[rgb]{0,0,0}\put(-149,-1036){\line( 1, 0){1050}}
}%
{\color[rgb]{0,0,0}\put(-149,-1111){\line( 1, 0){1050}}
}%
{\color[rgb]{0,0,0}\put(-149,-1186){\line( 1, 0){1050}}
}%
{\color[rgb]{0,0,0}\put(-149,-886){\line( 1, 0){1050}}
}%
{\color[rgb]{0,0,0}\put(-149,-811){\line( 1, 0){1050}}
}%
{\color[rgb]{0,0,0}\put(1501,-961){\line( 1, 0){1050}}
}%
{\color[rgb]{0,0,0}\put(1501,-1036){\line( 1, 0){1050}}
}%
{\color[rgb]{0,0,0}\put(1501,-1111){\line( 1, 0){1050}}
}%
{\color[rgb]{0,0,0}\put(1501,-1186){\line( 1, 0){1050}}
}%
{\color[rgb]{0,0,0}\put(1501,-886){\line( 1, 0){1050}}
}%
{\color[rgb]{0,0,0}\put(1501,-811){\line( 1, 0){1050}}
}%
{\color[rgb]{0,0,0}\put(1201,239){\circle*{76}}
}%
{\color[rgb]{0,0,0}\put(2851,239){\circle*{76}}
}%
{\color[rgb]{0,0,0}\put(3151,-961){\line( 1, 0){600}}
}%
{\color[rgb]{0,0,0}\put(3151,-1036){\line( 1, 0){600}}
}%
{\color[rgb]{0,0,0}\put(3151,-1111){\line( 1, 0){600}}
}%
{\color[rgb]{0,0,0}\put(3151,-1186){\line( 1, 0){600}}
}%
{\color[rgb]{0,0,0}\put(3151,-886){\line( 1, 0){600}}
}%
{\color[rgb]{0,0,0}\put(3151,-811){\line( 1, 0){600}}
}%
{\color[rgb]{0,0,0}\put(-449,-511){\line( 1, 0){1350}}
}%
{\color[rgb]{0,0,0}\put(-449,-211){\line( 1, 0){1350}}
}%
{\color[rgb]{0,0,0}\put(-449,-286){\line( 1, 0){1350}}
}%
{\color[rgb]{0,0,0}\put(-449,-361){\line( 1, 0){1350}}
}%
{\color[rgb]{0,0,0}\put(-449,-436){\line( 1, 0){1350}}
}%
{\color[rgb]{0,0,0}\put(-449,-136){\line( 1, 0){1350}}
}%
{\color[rgb]{0,0,0}\put(1501,-511){\line( 1, 0){1050}}
}%
{\color[rgb]{0,0,0}\put(1501,-211){\line( 1, 0){1050}}
}%
{\color[rgb]{0,0,0}\put(1501,-286){\line( 1, 0){1050}}
}%
{\color[rgb]{0,0,0}\put(1501,-361){\line( 1, 0){1050}}
}%
{\color[rgb]{0,0,0}\put(1501,-436){\line( 1, 0){1050}}
}%
{\color[rgb]{0,0,0}\put(1501,-136){\line( 1, 0){1050}}
}%
{\color[rgb]{0,0,0}\put(3151,-286){\line( 1, 0){900}}
}%
{\color[rgb]{0,0,0}\put(3151,-361){\line( 1, 0){900}}
}%
{\color[rgb]{0,0,0}\put(3151,-436){\line( 1, 0){900}}
}%
{\color[rgb]{0,0,0}\put(3151,-511){\line( 1, 0){900}}
}%
{\color[rgb]{0,0,0}\put(3151,-211){\line( 1, 0){900}}
}%
{\color[rgb]{0,0,0}\put(3151,-136){\line( 1, 0){900}}
}%
{\color[rgb]{0,0,0}\put(3451,239){\circle*{76}}
}%
{\color[rgb]{0,0,0}\put(3451,-1411){\circle{76}}
}%
{\color[rgb]{0,0,0}\put(901,-1261){\framebox(600,1200){swap}}
}%
{\color[rgb]{0,0,0}\put(1201,239){\line( 0,-1){300}}
}%
{\color[rgb]{0,0,0}\put(2551,-1261){\framebox(600,1200){$U$}}
}%
{\color[rgb]{0,0,0}\put(1801, 14){\framebox(450,450){$X$}}
}%
{\color[rgb]{0,0,0}\put(601,239){\line( 1, 0){1200}}
}%
{\color[rgb]{0,0,0}\put(2851,-61){\line( 0, 1){300}}
}%
{\color[rgb]{0,0,0}\put(2251,239){\line( 1, 0){1800}}
}%
{\color[rgb]{0,0,0}\put(151, 14){\framebox(450,450){$H$}}
}%
{\color[rgb]{0,0,0}\put(-149,239){\line( 1, 0){300}}
}%
{\color[rgb]{0,0,0}\put(-149,-1411){\line( 1, 0){3900}}
}%
{\color[rgb]{0,0,0}\put(3751,-811){\vector( 0,-1){825}}
}%
{\color[rgb]{0,0,0}\put(3451,239){\line( 0,-1){1688}}
}%
\put(-449,-1486){\makebox(0,0)[lb]{\smash{{\SetFigFont{12}{14.4}{\rmdefault}{\mddefault}{\updefault}{\color[rgb]{0,0,0}$\ket 0$}%
}}}}
\put(-449,164){\makebox(0,0)[lb]{\smash{{\SetFigFont{12}{14.4}{\rmdefault}{\mddefault}{\updefault}{\color[rgb]{0,0,0}$\ket 0$}%
}}}}
{\color[rgb]{0,0,0}\put(-149,-961){\line( 1, 0){1050}}
}%
\end{picture}%
 \end{center}
  \caption[Reduction to an antidegradable channel]{The antidegradable channel
    $\Psi$ constructed from $\Phi$.}
  \label{fig-adeg-reduction}
\end{figure}

Let $\Phi$ implement the transformation given by
\[ \Phi(\rho) = \ptr{E} U (\rho \tprod \ket 0 \bra 0) U^*, \] 
where, as before, the circuit for $\Phi$ can be assumed to be in this
form by introducing any ancillary qubits first and delaying the
tracing out of any qubits to the end of the circuit.
The antidegradable channel $\Psi$ will be constructed as
\begin{equation}\label{eqn-adeg-const}
  \Psi(\rho) = \frac{1}{2} \ket 0 \bra 0 \tprod \ket 0 \bra 0 +
  \frac{1}{2} \ket 1 \bra 1 \tprod \Phi(\rho).
\end{equation}
This is just the channel that applies $\Phi$ with probability one-half,
outputs $\ket 0$ with probability one-half, and outputs a flag qubit
in the space $\mathcal{C}$ to indicate which case has occurred.  In a
way very similar to the construction in
Section~\ref{scn-degr-sim}, this channel can be implemented using
a controlled-$U$ operation.  In this case, however, we will also need
the operation that swaps the states in two spaces (i.e.\ $\mathrm{swap}( \ket a
\ket b) = \ket b \ket a$).  An implementation of the channel $\Psi$ is
shown in Figure~\ref{fig-adeg-reduction}.
This circuit will, depending on the value of the control qubit in the
space $\mathcal{C}$ either apply $\Phi$ or output the pure state
$\ket 0$.

\begin{figure}
  \begin{center}
\setlength{\unitlength}{3947sp}%
\begingroup\makeatletter\ifx\SetFigFont\undefined%
\gdef\SetFigFont#1#2#3#4#5{%
  \reset@font\fontsize{#1}{#2pt}%
  \fontfamily{#3}\fontseries{#4}\fontshape{#5}%
  \selectfont}%
\fi\endgroup%
\begin{picture}(3477,1899)(286,-1423)
\put(301,-1036){\makebox(0,0)[lb]{\smash{{\SetFigFont{12}{14.4}{\rmdefault}{\mddefault}{\updefault}{\color[rgb]{0,0,0}$\ket 0$}%
}}}}
\thinlines
{\color[rgb]{0,0,0}\put(601,-1036){\line( 1, 0){300}}
}%
{\color[rgb]{0,0,0}\put(601,-1111){\line( 1, 0){300}}
}%
{\color[rgb]{0,0,0}\put(601,-1186){\line( 1, 0){300}}
}%
{\color[rgb]{0,0,0}\put(601,-886){\line( 1, 0){300}}
}%
{\color[rgb]{0,0,0}\put(601,-811){\line( 1, 0){300}}
}%
{\color[rgb]{0,0,0}\put(1501,-961){\line( 1, 0){1050}}
}%
{\color[rgb]{0,0,0}\put(1501,-1036){\line( 1, 0){1050}}
}%
{\color[rgb]{0,0,0}\put(1501,-1111){\line( 1, 0){1050}}
}%
{\color[rgb]{0,0,0}\put(1501,-1186){\line( 1, 0){1050}}
}%
{\color[rgb]{0,0,0}\put(1501,-886){\line( 1, 0){1050}}
}%
{\color[rgb]{0,0,0}\put(1501,-811){\line( 1, 0){1050}}
}%
{\color[rgb]{0,0,0}\put(1201,239){\circle*{76}}
}%
{\color[rgb]{0,0,0}\put(2851,239){\circle*{76}}
}%
{\color[rgb]{0,0,0}\put(3151,-961){\line( 1, 0){300}}
}%
{\color[rgb]{0,0,0}\put(3151,-1036){\line( 1, 0){300}}
}%
{\color[rgb]{0,0,0}\put(3151,-1111){\line( 1, 0){300}}
}%
{\color[rgb]{0,0,0}\put(3151,-1186){\line( 1, 0){300}}
}%
{\color[rgb]{0,0,0}\put(3151,-886){\line( 1, 0){300}}
}%
{\color[rgb]{0,0,0}\put(3151,-811){\line( 1, 0){300}}
}%
{\color[rgb]{0,0,0}\put(301,-511){\line( 1, 0){600}}
}%
{\color[rgb]{0,0,0}\put(301,-211){\line( 1, 0){600}}
}%
{\color[rgb]{0,0,0}\put(301,-286){\line( 1, 0){600}}
}%
{\color[rgb]{0,0,0}\put(301,-361){\line( 1, 0){600}}
}%
{\color[rgb]{0,0,0}\put(301,-436){\line( 1, 0){600}}
}%
{\color[rgb]{0,0,0}\put(301,-136){\line( 1, 0){600}}
}%
{\color[rgb]{0,0,0}\put(1501,-511){\line( 1, 0){1050}}
}%
{\color[rgb]{0,0,0}\put(1501,-211){\line( 1, 0){1050}}
}%
{\color[rgb]{0,0,0}\put(1501,-286){\line( 1, 0){1050}}
}%
{\color[rgb]{0,0,0}\put(1501,-361){\line( 1, 0){1050}}
}%
{\color[rgb]{0,0,0}\put(1501,-436){\line( 1, 0){1050}}
}%
{\color[rgb]{0,0,0}\put(1501,-136){\line( 1, 0){1050}}
}%
{\color[rgb]{0,0,0}\put(3151,-286){\line( 1, 0){600}}
}%
{\color[rgb]{0,0,0}\put(3151,-361){\line( 1, 0){600}}
}%
{\color[rgb]{0,0,0}\put(3151,-436){\line( 1, 0){600}}
}%
{\color[rgb]{0,0,0}\put(3151,-511){\line( 1, 0){600}}
}%
{\color[rgb]{0,0,0}\put(3151,-211){\line( 1, 0){600}}
}%
{\color[rgb]{0,0,0}\put(3151,-136){\line( 1, 0){600}}
}%
{\color[rgb]{0,0,0}\put(901,-1261){\framebox(600,1200){swap}}
}%
{\color[rgb]{0,0,0}\put(1201,239){\line( 0,-1){300}}
}%
{\color[rgb]{0,0,0}\put(2551,-1261){\framebox(600,1200){$U$}}
}%
{\color[rgb]{0,0,0}\put(1801, 14){\framebox(450,450){$X$}}
}%
{\color[rgb]{0,0,0}\put(301,239){\line( 1, 0){1500}}
}%
{\color[rgb]{0,0,0}\put(2851,-61){\line( 0, 1){300}}
}%
{\color[rgb]{0,0,0}\put(2251,239){\line( 1, 0){1500}}
}%
{\color[rgb]{0,0,0}\put(3451,-811){\vector( 0,-1){600}}
}%
{\color[rgb]{0,0,0}\put(601,-961){\line( 1, 0){300}}
}%
\end{picture}%
 \end{center}
  \caption[Anti-degrading channel for the channel in
  Figure~\ref{fig-adeg-reduction}]{The anti-degrading channel
    corresponding to the channel in $\Psi$ in
    Figure~\ref{fig-adeg-reduction}.}
  \label{fig-adeg-adegrading}
\end{figure}
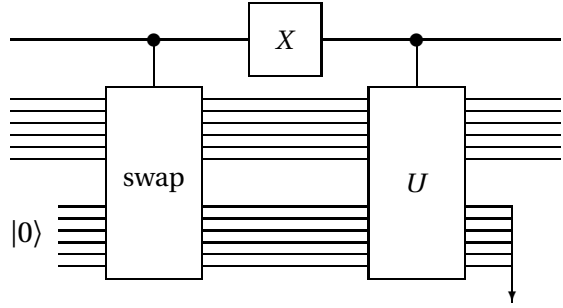

To show that the circuit $\Psi$ implements an antidegradable channel, we
explicitly construct the map $A_\Psi$ that maps the environment state of
$\Psi$ to the output state.  The environment state of $\Psi$ is once again
simply the state produced by $\Psi^C$, the complementary channel to
$\Psi$.  This channel is given by
\begin{equation}\label{eqn-adeg-compl}
  \Psi^C(\rho) = \frac{1}{2} \ket 0 \bra 0 \tprod \rho + \frac{1}{2} \ket 1 \bra 1 \tprod \Phi^C(\rho),
\end{equation}
where as before the channel $\Phi^C$ is given by
\[ \Phi^C(\rho) = \ptr{A} U ( \rho \tprod \ket 0 \bra 0) U^*. \] 
Given the state in Equation~\eqref{eqn-adeg-compl} it is not hard
to see how to map it to the state in Equation~\eqref{eqn-adeg-const}.
This can be done by implementing one of two operations, depending on
the value of the flag qubit in the space $\mathcal{C'}$, which is the
`copy' of the control qubit traced out in Figure~\ref{fig-adeg-reduction}.
If this qubit is in the state $\ket 0$, then the remainder of the
input state is $\rho$, the original input to $\Psi$, so that applying the
circuit for $\Phi$ produces the state $\Phi(\rho)$.  If the control qubit is
in the $\ket 1$ state, however, the remainder of the input state is
$\Phi^C(\rho)$.  This state can be discarded (i.e.\ traced out) and
ancillary qubits in the state $\ket 0$ can be swapped into the output
space.  
As before, the value of the qubit in $\mathcal{C'}$
needs to be flipped with a Pauli $X$ gate so that the state is exactly
correct.  A circuit implementing this is found in
Figure~\ref{fig-adeg-adegrading}.

To see that $A_\Psi$ correctly implements the anti-degrading map for $\Psi$, we compute
\begin{align*}
  A_\Psi( \Psi^C(\rho) )
  &= \frac{1}{2} A_\Psi\left(\ket 0 \bra 0 \tprod \rho + \ket 1 \bra 1 \tprod \Phi^C(\rho)\right) \\
  &= \frac{1}{2} \ket 1 \bra 1 \tprod \Phi(\rho) + \frac{1}{2} \ket 0 \bra 0 \tprod \ket 0 \bra 0 \\
  &= \Psi(\rho),
\end{align*}
where the final equality is Equation~\ref{eqn-adeg-const}.  This
demonstrates that the channel $\Psi$ constructed from $\Phi$ is
antidegradable.  In the following section the implications of this
construction for the hardness of distinguishing
antidegradable channels are considered.

\section{Distinguishing degradable and antidegradable channels}\label{scn-dist}

In this section we consider the implications of the constructions in
Sections~\ref{scn-degr-sim} and~\ref{scn-adegr-sim} for the
computational problem of distinguishing quantum channels.  These
constructions essentially embed any channel into either a degradable
or antidegradable channel.  This can be used to show that
distinguishing these channels is no easier than distinguishing general
channels.

To observe this in a more formal setting, we introduce the
problem of distinguishing two quantum
channels when they are provided as quantum circuits.  The
diamond norm $\dnorm{\Phi_1 - \Phi_2}$ determines the maximum
probability that an unknown channel $\Phi \in \{\Phi_1, \Phi_2\}$ can
be identified with a single use.  For this reason, we may formalize
the computational distinguishability problem in terms of evaluating
the diamond norm of the difference of two known channels.
\begin{problem}[Quantum Circuit Distinguishability]\label{prob-qcd}
  For constants $0 \leq b < a \leq 2$, the input consists of
  quantum mixed-state circuits $\Phi_1$ and $\Phi_2$ that implement transformations in
  $\transform{A,A}$.
  The promise problem is to distinguish the two cases:
  \begin{description}
    \item[Yes:] $\dnorm{\Phi_1 - \Phi_2} \geq a$,
    \item[No:] $\dnorm{\Phi_1 - \Phi_2} \leq b$.
  \end{description}
\end{problem}
\noindent This problem is introduced and shown to be
\class{PSPACE}-complete for all $0 < b < a < 2$ in~\cite{RosgenW05}.
The distinguishability problem as originally defined allows the input
and output dimensions of the channels to differ, but as discussed in Section~\ref{scn-sim}
this can be avoided by padding the smaller of the two spaces.
For conciseness, this problem will be abbreviated $\prob{QCD}_{a,b}$.
Restricting the channels $\Phi_1, \Phi_2$ in Problem~\ref{prob-qcd} to
degradable channels results in the problem $\prob{Degradable
  QCD}_{a,b}$, whereas the restriction to antidegradable channels
gives the problem $\prob{Antidegradable QCD}_{a,b}$.  The main result
of this paper is that these problems remain \class{PSPACE}-complete.

To show the hardness of these restricted distinguishability problems,
we show that the constructions of Sections~\ref{scn-degr-sim}
and~\ref{scn-adegr-sim} reduce the general \prob{QCD} problem to these
two restricted problems.  These two constructions can be efficiently
implemented when the input channels are given as quantum circuits:
this can be seen from the circuit representations in
Figures~\ref{fig-degrad-reduction} and~\ref{fig-adeg-reduction}.  
It will be shown that these reductions prove
the hardness of the restricted distinguishability problems,
which suffices to prove that they are \class{PSPACE}-complete: this is
because they are contained in \class{PSPACE} by the same algorithm
used for the general \prob{QCD} problem, which can be found
in~\cite{RosgenW05}.

The primary ingredient in the proof that these constructions reduce
the distinguishability problem to degradable and antidegradable
channels is a proof that when applied to each of a pair of channels,
the constructions preserve the diamond norm of the difference of the
two channels.
This is not difficult to see from the
output of the constructions, given by
Equations~\eqref{eqn-degr-const} and~\eqref{eqn-adeg-const}, 
but for completeness this is argued formally in the following lemma.

\begin{lemma}\label{lem-dnorm}
  Let $\Phi_1, \Phi_2$ be quantum circuits implementing
  transformations in $\transform{A,A}$.  If the channels $\Psi_1, \Psi_2 \in \transform{A, C \tprod A}$
  are given by
  \begin{equation*}
    \Psi_i(\rho) 
    = \frac{1}{2} \ket 0 \bra 0 \tprod \rho 
    + \frac{1}{2} \ket 1 \bra 1 \tprod \Phi_i(\rho),
  \end{equation*}
  and the channels $\Lambda_1, \Lambda_2 \in \transform{A, C \tprod A}$ are
  given by
  \begin{equation*} 
    \Lambda_i(\rho) 
    = \frac{1}{2} \ket 0 \bra 0 \tprod \ket 0 \bra 0
    + \frac{1}{2} \ket 1 \bra 1 \tprod \Phi_i(\rho),
  \end{equation*}
  for $i \in \{1,2\}$, then
  \begin{equation*}
    \dnorm{\Psi_1 - \Psi_2} 
    = \dnorm{\Lambda_1 - \Lambda_2} 
    = \frac{1}{2} \dnorm{\Phi_1 - \Phi_2}.
  \end{equation*}
  \begin{proof}
    Let $\rho \in \density{A \tprod F}$ be an arbitrary state.  Then
    \begin{align*}
      \tnorm{(\Psi_1 \tprod \tidentity{F} - \Psi_2 \tprod \tidentity{F})(\rho)}
      &= \frac{1}{2} \tnorm{ 
           \ket 0 \bra 0 \tprod (\rho - \rho) 
        + \ket 1 \bra 1 \tprod ( 
            \left[ \Phi_1 \tprod \tidentity{F} - \Phi_2 \tprod \tidentity{F}
            \right] (\rho)) } \\
      &= \frac{1}{2} \tnorm{ 
       \left( \Phi_1 \tprod \tidentity{F} - \Phi_2 \tprod \tidentity{F} \right) (\rho) }.
    \end{align*}
    Since the diamond norm may be defined as the maximization over all
    such states $\rho$, this implies the equality $\dnorm{\Psi_1 -
      \Psi_2} = \dnorm{\Phi_1 - \Phi_2}/2$.  The same argument implies
    that $\dnorm{\Lambda_1 - \Lambda_2} = \dnorm{\Phi_1 - \Phi_2}/2$.
  \end{proof}
\end{lemma}

This lemma implies that distinguishing degradable or antidegradable
channels is \class{PSPACE}-complete for all $0 < b < a < 1$, using
the hardness result known for the general problem, $\prob{QCD}_{2a,
  2b}$~\cite{RosgenW05}.  Notice that we lose a factor of two in the
parameters $a$ and $b$: this is because the diamond norm of the
constructed channels is half the norm of the original channels.

This result can be strengthened with parallel repetition.
The strategy is to take an
instance $(\Psi_1, \Psi_2)$ of $\prob{Degradable QCD}_{1-\varepsilon,\varepsilon}$ and
construct the instance $(\Psi_1^{\tensor k}, \Psi_2^{\tensor k})$.  This
second instance will have outputs that are more distinguishable, for
the simple reason that there are more copies of the states to be
distinguished available.  This will send the norm for `yes' instances
of the problem from $1-\varepsilon$ to a value close to 2, but it also has the
property that the norm of `no' instances is not made too large.  This
is a straightforward consequence of the bounds in
Lemma~\ref{lem-direct-product}, which appears in~\cite{RosgenW05} as
part of an efficient procedure for polarizing the diamond norm.  This
technique can be applied to both \prob{Degradable QCD} and
\prob{Antidegradable QCD} as the classes of degradable and
antidegradable channels are closed under parallel repetition.

\begin{theorem}\label{thm-degr-hard}
  For any choice of constants $0 < b < a < 2$, both of the problems
  $\prob{Degradable QCD}_{a,b}$ and $\prob{Antidegradable QCD}_{a,b}$ are \class{PSPACE}-complete.

  \begin{proof}
    These problems are in \class{PSPACE} as they are restrictions of the
    general \prob{QCD} problem~\cite{RosgenW05}.
    To see that it they are also
    \class{PSPACE}-hard, take an instance $(\Phi_1, \Phi_2)$ of the
    $\prob{QCD}_{2-2\varepsilon,2\varepsilon}$, for $\varepsilon>0$ a small
    constant.  We will reduce this problem to \prob{Degradable QCD}.

    Applying the construction of Section~\ref{scn-degr-sim}
    to $(\Phi_1, \Phi_2)$ results in a pair of circuits $(\Psi_1,
    \Psi_2)$ that form an instance of
    $\prob{Degradable QCD}_{1-\varepsilon,\varepsilon}$, by
    Lemma~\ref{lem-dnorm}.  
    As the degradable channels are closed under tensor products, 
    $(\Psi_1^{\tprod k}, \Psi_2^{\tprod k})$ gives a pair of circuits
    implementing degradable channels.  By
    Lemma~\ref{lem-direct-product}, we have
    \begin{align*}
      \dnorm{\Psi_1 - \Psi_2} \geq 1 - \varepsilon & \implies
      \dnorm{\Psi_1^{\tprod k} - \Psi_2^{\tprod k}} \geq 2 - 2 e^{-k (1 - 2 \varepsilon)/8}, \\
      \dnorm{\Psi_1 - \Psi_2} \leq \varepsilon & \implies
      \dnorm{\Psi_1^{\tprod k} - \Psi_2^{\tprod k}} \leq k \varepsilon. 
    \end{align*}
    Then, for any $0 < b < a < 2$, choosing $k \geq -16 \ln (1 - a/2)$
    and $\varepsilon \leq \min \{ 1/4, b / k \}$ implies the desired inequalities
    $2 - 2 e^{-k(1 - 2 \varepsilon)/8} > a$ and
    $k   \varepsilon < b$.
    This shows the \class{PSPACE} hardness of
    $\prob{Degradable QCD}_{a,b}$.  The case of $\prob{Antidegradable
      QCD}_{a,b}$ is completely symmetric, with the exception that we
    use the construction of Section~\ref{scn-adegr-sim} to obtain
    antidegradable channels $\Psi_1$ and $\Psi_2$.
  \end{proof}
\end{theorem}

\section{Conclusion}

This paper has presented a construction for embedding an arbitrary
channel into a degradable channel due to Cubitt, Ruskai, and
Smith~\cite{CubittR+08}, as well as a closely related construction for
antidegradable channels.  These constructions can be efficiently
implemented on quantum circuits, so that instances of the quantum
circuit distinguishability problem can be mapped to degradable or
antidegradable channels.
The main result is that the distinguishability problem
on quantum circuits remains hard when restricted to either the class
of degradable channels or the class of antidegradable channels.

\section*{Acknowledgements}

I am grateful for discussions with John Watrous, as well as helpful
comments from Richard Cleve, Achim Kempf, and Steve Fenner.
This work has been supported by the Centre for Quantum Technologies at
the National University of Singapore, as well the Bell Family Fund for
Quantum Computing, while the author was a student at the Institute for
Quantum Computing at the University of Waterloo.

\newcommand{\arxiv}[2][quant-ph]{\href{http://arxiv.org/abs/#2}{arXiv:#2 [#1]}}
  \newcommand{\oldarxiv}[2][quant-ph]{\href{http://arxiv.org/abs/#1/#2}{arXiv:%
#1/#2}}

\end{document}